 \documentclass[12pt]{article}
 \usepackage{amsmath,amsfonts,latexsym,amssymb,amsthm,mathrsfs,}
 \usepackage{times}

 \newtheorem{thm}{Theorem}[section]

 \newtheorem{prop}[thm]{Proposition}
 \theoremstyle{definition}
 \newtheorem{defn}[thm]{Definition}
 \theoremstyle{remark}

 \numberwithin{equation}{section}

\begin{document}
 
\title{Wave equations with \\ non-commutative space and time}
\author{ {\sf Rainer Verch}\footnote{email: rainer.verch@uni-leipzig.de} \\[12pt]
\normalsize
Institut f\"ur Theoretische Physik\\
\normalsize Universit\"at Leipzig,
D-04009 Leipzig,
Germany}
\date{ }
\maketitle

\begin{abstract}
${}$ \quad  \quad The behaviour of solutions to the partial differential equation \linebreak[4]
$(D + \lambda W)f_\lambda = 0$ is
discussed, where $D$ is a normal hyperbolic partial differential operator, or pre-normal hyperbolic
operator, on $n$-dimensional Minkowski spacetime. The potential term $W$ is a $C_0^\infty$ kernel operator
which, in general, will be non-local in time, and $\lambda$ is a complex parameter. A result is presented
which states that there are unique advanced and retarded Green's operators for this partial differential
equation if $|\lambda|$ is small enough (and also for a larger set of $\lambda$ values). Moreover, a 
scattering operator can be defined if the $\lambda$ values admit advanced and retarded Green operators.
In general, however, the Cauchy-problem will be ill-posed, and examples will be given to that effect. 
It will also be explained that potential terms arising from non-commutative products on function spaces can
be approximated by $C_0^\infty$ kernel operators and that, thereby, scattering by a non-commutative potential
can be investigated, also when the solution spaces are (2nd) quantized. Furthermore, a discussion will
be given which links the scattering transformations, which thereby arise from non-commutative potentials, to
observables of quantum fields on non-commutative spacetimes through
``Bogoliubov's formula''. In particular, this helps to shed light on the
question how observables arise for quantum fields on Lorentzian spectral geometries.
\end{abstract}

\maketitle

\section{Introduction}

The present contribution essentially reports on the results of a recent article by Gandalf Lechner and the
present author \cite{LechnerVerch} to which --- often without explicit mentioning --- the reader is
referred for considerable further details and discussion. The investigation of the said article is embedded
in the quest for understanding the relevant structures of quantum field theories on non-commutative spacetimes.
There are various ways of approaching this theme, and sometimes the various strands don't seem to connect
very well. An approach advocated by the present author \cite{Verch:2011,PaschkeVerch,BorrisVerch}
is to attempt and combine a general description 
of non-commutative spacetimes in a framework of Lorentzian spectral geometry
\cite{Strohmaier,Franco-Thesis,vdDPR,PaschkeVerch,Verch:2011} with the
basic principles of local covariant quantum field theory \cite{BFV,FewsterVerch:2012}. 
While this undertaking has some promising aspects
from a structural point of view,
it is hampered by being very ambitious --- perhaps, overly ambitious --- from the outset, in that it attempts
to design a framework for quantum field theory on a large class of non-commutative spacetimes, despite the fact
that it is not even clear if we know what ``the general structure'' of quantum field theory on a fixed
non-commutative spacetime should be. In particular, it is unclear if ``the'' is an appropriate prefix, as
there could be many structurally quite different concepts of quantum field theory on a given non-commutative
spacetime which nevertheless appear reasonable for the --- hypothetical --- physical situation for which they
are conceived. Naturally, the idea is to bring a suitably generalized principle of local covariance into play 
so as to narrow down the potentially vast variety of choices one could make. Judging from the present status of
the development of quantum field theory on non-commutative spacetimes, there seems to lie still a long way ahead of
us before this could be achieved.

Yet, we will sketch elements of our preferred approach here. The starting point is to
set down some rudimentary conditions for a Lorentzian spectral triple, a sort of nucleus for Lorentzian
spectral geometry, and to notice that, irrespective of several details (mostly pertaining to the
analytical structure of a Lorentzian spectral triple), one can associate a $C^*$-algebra of
canonical anti-commutation relations to any Lorentzian spectral triple. This amounts to an abstract
construction of the quantized Dirac field on the --- potentially non-commutative --- spacetime geometry
described by the Lorentzian spectral triple. The usual Minkowski spacetime as well as Moyal deformed 
Minkowski spacetime will serve as examples. In both examples, one obtains the identical $C^*$ algebra
of the quantized Dirac field, and thus there arises the question of how to obtain the information
that in the first case one should regard the $C^*$ algebra as describing the quantized Dirac field on Minkowski
spacetime, while in the second case as describing the quantized Dirac field on Moyal deformed Minkowski
spacetime. As we will discuss to some extent in Section 2, this can be achieved by looking at the action
of the algebra $\mathcal{A}$ of the Lorentzian spectral triple on the quantum field operators in the
respective cases. This, in turn, may be derived from scattering operators which relate solutions $f_\lambda$ to 
the Dirac equation with an ``external interaction potential'',
$$ (D + \lambda V_a) f_\lambda = 0 $$
to solutions $f_0$ of the ``free'' Dirac equation $D f_0 = 0$, at asymptotically early or late times, akin to
potential scattering in quantum mechanics. Here, $D$ is the Dirac operator on Minkowski spacetime, and 
$V_a$ is an action of an element $a$ of $\mathcal{A}$ on the $f_\lambda$ which are viewed as elements of the
Lorentzian spectral triple's Hilbert space, $\mathcal{H}$. Thus, in the case of Minkowski spacetime, 
$a$ will typically be a (real-valued) Schwartz function on $\mathbb{R}^4$ and $V_a$ will then amount to
pointwise multiplication of $f_\lambda$ by $a$, $(V_a f_\lambda)(x) = a(x)f_\lambda(x)$ $(x \in \mathbb{R}^4)$.
In contrast, in the case of Moyal deformed Minkowski spacetime, $V_a$ is given by the Moyal product of
$a$ and $f_\lambda$ (see Section 2),
$$ (V_a f_\lambda)(x) = (a \star f_\lambda) (x) \quad \ (x \in \mathbb{R}^4) \,.$$
As will be explained in Sec.\ 2, the scattering operators obtained from the potential scattering in both
situations lead to Bogoliubov transformations $\alpha_\lambda$ on the $C^*$ algebra of the quantized
Dirac field for any $a$, and differentiation with
respect to the coupling strength parameter $\lambda$ induces operators $X(a)$ (in the Hilbert space of
the vacuum representation of the ``free'' quantized, massless Dirac field on Minkowski spacetime) such that
$$ \left. \frac{d}{d\lambda} \right|_{\lambda} \,\alpha_\lambda(\Psi(h)) = i[X(a),\Psi(h)] \,.$$
Here, $\Psi(h)$ is a quantized Dirac field operator, and following the line of thought of 
``Bogoliubov's formula'', it may be regarded as an obervable of the quantized Dirac field --- in this case,
linearly dependent on elements $a$ in the algebra $\mathcal{A}$ of the underlying Lorentzian spectral triple.
It is the assignment $a \mapsto X(a)$ and the algebraic relations of the $X(a)$ for various $a$ which
encode the information that the quantized Dirac field propagates in one case on usual Minkowski spacetime,
or in the other case, on Moyal deformed Minkowski spacetime. This conceptual framework can in principle
be transferred to more general Lorentzian spectral triples.

At any rate, in order to obtain the assignment $a \mapsto X(a)$, the first step is to obtain scattering
operators for a Dirac equation of the form
$$  D f_\lambda + \lambda a \star f_\lambda = 0 \,.$$
The problem here is that the Moyal product acts non-locally, also with respect to any time-direction
on Minkowski spacetime, and therefore one cannot put this equation into the form of a first-order
system. However, the operators $V_a f_\lambda = a \star f_\lambda $ can be approximated by operators
\begin{align} \label{eqn:Def-W}
Wh(x) = \int w(x,y) h(y) d^4 y 
\end{align}
where $w$ is a $C_0^\infty$ kernel (matrix-valued, since $h$ has several components). In \cite{LechnerVerch},
we have investigated question of existence and uniqueness of solutions $f_\lambda$ to 
the equation
\begin{align} \label{eqn:D+W}
(D + \lambda W) f_\lambda = 0
\end{align}
where $D$ is either a 2nd order normal hyperbolic partial differential operator, or a pre-hyperbolic 
partial differential operator (meaning that $D$ is first order, and there is another first order operator
such that $DD'$ and $D'D$ are normal hyperbolic --- the Dirac operator is an example)
on $n$-dimensional Minkowski spacetime, and where $W$ is given by a $C_0^\infty$ kernel as in 
\eqref{eqn:Def-W}. The results of \cite{LechnerVerch}, which will be summarized in more detail in Section 3,
are as follows: If $|\lambda|$ is sufficiently small, then there are unique advanced and retarded
fundamental solutions (Green operators) for \eqref{eqn:D+W} (and in fact, the Green operators are 
meromorphic in $\lambda$); in general, the Cauchy-problem for \eqref{eqn:D+W} is ill-posed (admitting arbitrary 
$W$); nevertheless, scattering operators for \eqref{eqn:D+W} can be uniquely constructed.
We elaborate a bit more on the perspectives of these results for obtaining the operators
$X(a)$ mentioned above in the final Section 4.

\section{Quantum fields on Lorentzian spectral triples}

 In the spectral geometry approach to non-commutative
spaces, the description of a commutative or non-commutative manifold is given in terms of
a {\em spectral triple} $(\mathcal{A},D,\mathcal{H})$, where $\mathcal{H}$ is a Hilbert space,
$\mathcal{A}$ is a $*$-algebra of operators acting in $\mathcal{H}$, and $D$ is a distinguished
(unbounded) operator on a suitable domain in $\mathcal{H}$. In the case 
where the spectral triple corresponds to a ``commutative'' compact Riemannian
manifold with spin structure, $\mathcal{H}$ is formed by the space of $L^2$ spinor fields on the manifold, $\mathcal{A}$
is the --- commutative --- $*$-algebra of complex valued functions on the manifold, and 
$D$ is the Dirac operator. On the other hand, Connes' reconstruction theorem \cite{Connes3} shows that,
if a spectral triple has a commutative algebra $\mathcal{A}$, then it actually arises from a 
compact Riemannian manifold with spin structure in the way just indicated. That however needs further
data objects for a spectral triple and relations among them and, in particular, relations of the
additional data objects with the operator $D$. We refer to \cite{Connes,Connes3,GBVF01} for considerable further
discussion. At any rate, replacing the commutative algebra $\mathcal{A}$ by a non-commutative algebra (while
preserving relations between the data objects) leads to the concept of a non-commutative compact Riemannian
manifold in the spectral geometry approach. For examples, see \cite{GBVF01}. 

A physical spacetime is a four-dimensional Lorentzian spacetime and it is usually taken to be
globally hyperbolic, and hence non-compact, to avoid causal pathologies. This means that spacetimes
do not fit readily into the spectral geometry approach which, therefore, must be suitably generalized to
a form of Lorentzian spectral geometry. 
While such a generalization of spectral geometry doesn't appear to have reached
a final form up to now, there is certain progress in this direction 
\cite{Strohmaier,PaschkeVerch,vdDPR,Franco-Thesis,franc-eckst-2013,franc-eckst-2014}. The basic idea
is that a Lorentzian spectral geometry is again described by a spectral triple $(\mathcal{A},D,\mathcal{H},{\sf x}_i)$
but with further data objects ${\sf x}_i$ which are different from the compact Riemannian manifold case mentioned
before, and have different relations among each other, and with $D$. As mentioned, the dicussion has not
reached a final form as to what the ${\sf x}_i$ and their relations are. However, for any promising choice, it is expected
that the following holds: 
\\[4pt]
Suppose two (globally hyperbolic) Lorentzian spacetimes $\boldsymbol{M}$ and
$\tilde{\boldsymbol{M}}$ (with spin structures) are described by Lorentzian spectral triples
$(\mathcal{A},D,\mathcal{H},{\sf x}_i)$ and \linebreak[4]
$(\tilde{\mathcal{A}},\tilde{D},\tilde{\mathcal{H}},\tilde{{\sf x}}_i)$. Then
the two Lorentzian spectral triples are unitarily equivalent if and only if $\boldsymbol{M}$ and $\tilde{\boldsymbol{M}}$ are
isometric with equivalent spin structures. Here, the two Lorentzian spectral triples are called unitarily
equivalent if there is a unitary operator $U : \mathcal{H} \to \tilde{\mathcal{H}}$ such that
\begin{align} \label{eqn:uni-equiv}
 \tilde{\mathcal{A}} = U \mathcal{A} U^{-1}\,, \ \ \tilde{{\sf x}}_i = U {\sf x}_i U^{-1} \,, \ \
 [\,[\tilde{D},UaU^{-1}],\tilde{b}\,] = 
 [U [D,a] U^{-1},\tilde{b}] 
\end{align}
for all $a \in \mathcal{A}$, $\tilde{b} \in \tilde{\mathcal{A}}$,
where $[A,B] = AB -BA$.
\\[6pt]
Let us introduce the abbreviations $\boldsymbol{L} = (\mathcal{A},D,\mathcal{H},{\sf x}_i)$ for a 
Lorentzian spectral triple and $\boldsymbol{L} \overset{\psi_U}{\longrightarrow} \tilde{\boldsymbol{L}}$ for
a unitary equivalence morphism between spectral triples induced by a unitary $U$ as in \eqref{eqn:uni-equiv}.

At this point it is in order to briefly mention the basic structure of local covariant quantum field theory
(see \cite{BFV,FewsterVerch:2012} for further details not elaborated on in these writings).
A local covariant quantum field theory consists of an assignment $\boldsymbol{M} \to \mathscr{A}(\boldsymbol{M})$
of $*$-algebras (or $C^*$-algebras) to globally hyperbolic spacetimes $\boldsymbol{M}$. The $\mathscr{A}(\boldsymbol{M})$
are the algebras of observables, or more generally, of the quantum field (of a given type) on the spacetime 
$\boldsymbol{M}$. Additionally, whenever there is an isometric hyperbolic embedding $\boldsymbol{M} 
\overset{\psi}{\longrightarrow} \tilde{\boldsymbol{M}}$ --- i.e.\ if $\boldsymbol{M}$ can be viewed as a
globally hyperbolic sub-spacetime of $\tilde{\boldsymbol{M}}$ --- then there should be an injective
$*$-algebra morphism $\mathscr{A}(\boldsymbol{M}) \overset{\alpha_\psi}{\longrightarrow} \mathscr{A}(\tilde{\boldsymbol{M}})$;
moreover, the composition law $\alpha_{\psi_1 \circ \psi_2} = \alpha_{\psi_1} \circ \alpha_{\psi_2}$ is required
to hold. Expressed in more mathematical terms, this says that a local covariant quantum field theory is a 
functor from the category of globally hyperbolic spacetimes (all four-dimensional), with isometric hyperbolic 
embeddings as arrows, to the category of $*$-algebras, with monomorphisms as arrows. 
The interesting point is that this does not amount to just dressing up quantum field theory on general spacetime
manifolds in a fancy mathematical coat, but that it has led to significant new insights and results in quantum
field theory in curved spacetimes. We will not report on this issue here any further but just refer to the recent
survey \cite{FewsterVerch:2015} and literature cited there for a fuller discussion. 

It is suggestive to try and carry over this line of approach to quantum field theory on non-commutative spacetimes
essentially by replacing the category of globally hyperbolic spacetimes by the category of Lorentzian spectral triples
with unitary equivalences as arrows --- for a start (this is certainly not general enough; see remark to follow).
Then a ``covariant'' quantum field theory on Lorentzian non-commutative spacetimes should be given by an assignment
$\boldsymbol{L} \to \mathscr{A}(\boldsymbol{L})$ of a $*$-algebra $\mathscr{A}(\boldsymbol{L})$ to any
Lorentzian spectral triple $\boldsymbol{L}$ together with injective $*$-algebra morphisms $\mathscr{A}(\psi_U)$
for any unitary equivalence $\boldsymbol{L} \overset{\psi_U}{\longrightarrow} \tilde{\boldsymbol{L}}$ featuring
the functorial property $\mathscr{A}(\psi_{U_1}) \circ \mathscr{A}(\psi_{U_2}) = \mathscr{A}(\psi_{U_1 U_2})$.

The remark that this is not enough is in order now. One of the strengths of the local covariant framework
for quantum field theory stems from the fact that an embedding of a spacetime into a larger one is 
accompanied by an embedding of the corresponding quantum field theories. One would have to devise a similar
embedding for Lorentzian spectral triples which gives similarly rise to an embedding of the associated
quantum field theory. It is not clear what a suitable concept of embedding of Lorentzian spectral triples,
to this end, would amount to. A first working hypothesis might be to replace the unitary $U$ by a partial
isometry, or a partial isometry combined with a suitable generalization of the equalities in \eqref{eqn:uni-equiv}
as holding only up to ``neglible corrections'', an idea which is in fact of some importance in spectral
geometry \cite{Connes,GBVF01,PaschkeVerch}. At any rate, one encounters the problem of what replaces the concept of locality in quantum field
theory on non-commutative spacetimes at this point.

We are hopeful that our investigation reported on here will contribute to gaining further understanding of these
matters. To illustrate how we hope to approach that matter, we will be more concrete and consider a very simple
model for a non-commutative version of Minkowski spacetime, the Moyal-deformed Minkowski spacetime. As we have mentioned
before, it is not entirely clear what the assumptions on a Lorentzian spectral triple ultimately should be, but
in case they were set up in such a way that Moyal-deformed Minkowski spacetime does not fit into the framework,
then the list of examples of quantum field theories on non-commutative Lorentzian spectral triple spacetimes
would run thin indeed. Therefore, we anticipate that Moyal-deformed Minkowski spacetime (as well as Minkowski spacetime
as such) can be modelled as Lorentzian spectral triples. To that end, the primary indication and motivation comes from the discussion
in \cite{GGBI04} (see also \cite{BorrisVerch}) where Moyal planes are described as generalized Riemannian spectral triples. 
That description can be given a Lorentzian variant which one would expect to bear central features of Lorentzian
spectral geometry. We outline it here, very crudely. A Lorentzian spectral triple for ``commutative'' Minkowski spacetime
would start from taking a Hilbert space, $\mathcal{H}$, of $L^2$ spinors on Minkowski spacetime. There is no
Poincar\'e covariant notion of $L^2$ spinors on Minkowski spacetime but there are many possible choices depending on
a choice of time direction; consequently, the choice made will be recorded and forms a piece of data of the spectral geometry.
The algebra $\mathcal{A}$ can be taken to be $\mathscr{S}(\mathbb{R}^4)$, the Schwartz functions on Minkowski spacetime,
with their commutative pointwise multiplication as algebra product. The obvious choice for $D$ is the usual 
Lorentzian Dirac operator on Minkowski spacetime. To obtain a Lorentzian spectral triple for Moyal-deformed Minkowski
spacetime, one only needs to replace the commutative algebra $\mathscr{S}(\mathbb{R}^4)$ by 
$\mathscr{S}_\star(\mathbb{R}^4)$, the Schwartz functions with the non-commutative Moyal product \footnote{where
$(p \cdot z) = \sum_{\mu = 0,\ldots,3} p_\mu z_\mu$ is the Euclidean scalar product on $\mathbb{R}^4$ and $\theta$ is
a real invertible symplectic $4 \times 4$ matrix which is kept fixed (and usually chosen with ${\rm det}(\theta) = 1$)}
\begin{align} \label{MoyalProduct}
 f \star h = \int_{\mathbb{R}^4} d^4p \, \int_{\mathbb{R}^4} d^4z \, {\rm e}^{2\pi i (p \cdot z)} f(x - \theta p)h(x - z)
\end{align}
as algebra product, rendering a non-commutative algebra $\mathcal{A}$. 
(For more details, see the references \cite{Rieffel:1992,GGBI04,BuchholzLechnerSummers:2011} .)

The next step of interest to us is setting up a quantum field theory on Minkowski spacetime given in form
of a Lorentzian spectral triple. Since spinors and the Dirac operator appear in that description of Minkowski spacetime,
as a beginning step it appears most natural to start with the free quantized Dirac field. There is indeed a very
simple way of associating to the Lorentzian spectral triple 
$\boldsymbol{L}_0 = (\mathscr{S}(\mathbb{R}^4),D,\mathcal{H})$ of 
Minkowski space the quantized Dirac field: By defining $\mathscr{F}(\boldsymbol{L}_0)$ as the
CAR algebra --- algebra of canonical anti-commutation relations --- which is the unique $C^*$-algebra
generated by a unit element ${\bf 1}$ and by elements $\Psi(f)$, $f \in \mathcal{H}$, with the properties:
\begin{align*}
 (1) & \quad \ f \mapsto \Psi(f) \ \ \ \text{is linear}\,, \\
 (2) & \quad \ \Psi(D f) = 0\,, \\
 (3) & \quad \ \Psi(f)^* = \Psi(\Gamma f) \,, \\
 (4) & \quad \ \Psi(f)^*\Psi(h) + \Psi(h)\Psi(f)^* = i\langle f ,\gamma_0 R h \rangle {\bf 1} \,.
\end{align*}
Here, the $\Gamma$ appearing on the right hand side of (3) is a preferred complex
conjugation on $\mathcal{H}$ which is actually contained (previously unmentioned) in
the full list of data for a Lorentzian spectral triple. In (4), the notation is
$\langle f, h \rangle$ is  the scalar product of the
Hilbert space $\mathcal{H}$ on the right hand side and an operator $\gamma_0$ on $\mathcal{H}$
which is a further datum of the Lorentzian spectral triple carrying the information about
the time-direction that has been chosen to obtain a Lorentzian scalar product on the spinors.
Incidentally, in the example at hand, $\gamma_0$ coincides with the Dirac matrix $\gamma_0$ if
$\mathcal{H}$ is taken as $L^2(\mathbb{R}^4,\mathbb{C}^4)$ with scalar product
\begin{align} \label{eqn:scalarproduct}
\langle f,h \rangle  = \sum_{A = 0}^3 \int_{\mathbb{R}^4} \overline{f}_A(x) h_A(x)\, d^4x \,
\end{align}
Finally, $R = R^+ - R^-$ is the difference of retarded and advanced fundamental solutions
to the Dirac operator $D$ which are defined as (suitably continuous) linear operators
$R^\pm : \mathscr{S}(\mathbb{R}^4,\mathbb{C}^4) \to C^\infty(\mathbb{R}^4,\mathbb{C}^4)$
such that 
\begin{align} \label{eqn:Green-Op}
 DR^\pm f = R^\pm Df = f \ \ \text{and} \ \ {\rm supp}(R^\pm f) \subset J^\pm({\rm supp}(f))
\end{align}
where $J^\pm({\rm supp}(f))$ means causal future (+) / causal past $(-)$ of ${\rm supp}(f)$
in Minkowski spacetime. In this case, it turns out that $i \langle f , \gamma_0 R h \rangle$ endows
$\mathscr{S}(\mathbb{R}^4)$ with a semi-definite sesquilinear form, and the kernel of that form
coincides with the kernel of $R$.

This defines the algebra $\mathscr{F}(\boldsymbol{L}_0)$ of the Lorentzian spectral triple
of Minkowski spacetime and it actually coincides with the usual CAR algebra of the quantized
Dirac field in Minkowski spacetime. One observes that $\mathscr{F}(\boldsymbol{L}_0)$ is
nothing more than an abstract ``2nd quantization'' of the Hilbert space $\mathcal{H}$ (together with the
complex conjugation $\Gamma$) of $\boldsymbol{L}_0$. Moreover, one also readily observes that
the algebra $\mathcal{A} = \mathscr{S}(\mathbb{R}^4)$ appearing in $\boldsymbol{L}_0$
does not enter the construction of $\mathscr{F}(\boldsymbol{L}_0)$. On one hand, that can be taken
as an advantage since it allows it to directly generalize the construction of 
$\mathscr{F}(\boldsymbol{L}_0)$ from $\boldsymbol{L}_0$ to the case of 
$\boldsymbol{L}_\star = (\mathscr{S}_\star(\mathbb{R}^4),D,\mathcal{H})$. Clearly, this
results in $\mathscr{F}(\boldsymbol{L}_\star) = \mathscr{F}(\boldsymbol{L}_0)$ since, as mentioned,
the data of $\boldsymbol{L}_0$ and $\boldsymbol{L}_\star$ are identical apart from 
the different algebras $\mathscr{S}(\mathbb{R}^4)$ and $\mathscr{S}_\star(\mathbb{R}^4)$ which
however don't appear in the construction. On the other hand, that invokes the question
where the information is stored that $\mathscr{F}(\boldsymbol{L}_\star)$ is the algebra
of the quantized Dirac field on the non-commutative Moyal-deformed Minkowski spacetime whereas
$\mathscr{F}(\boldsymbol{L}_0)$ is the algebra of the the quantized Dirac field on the 
classical, ``commutative'' Minkowski spacetime. Obviously, in order to see the difference,
one needs some kind of action of the algebras $\mathscr{S}_\star(\mathbb{R}^4)$ and 
$\mathscr{S}(\mathbb{R}^4)$, respectively, on the algebra of the quantized Dirac field.

One possible such action can be derived from a scattering situation. Let $a$ be a (real)
test-function of Schwartz type on $\mathbb{R}^4$. Regarding $a$ as an element of 
the commutative algebra $\mathscr{S}(\mathbb{R}^4)$, one can modify the free Dirac equation
$Df_0 = 0$ to the Dirac equation
\begin{align} \label{eqn:Dirac_lambda}
(D + \lambda V_a)f_\lambda = 0 
\end{align}
with a potential term $V_a$ and small (real) parameter $\lambda$, where
\begin{align} \label{eqn:comm_potential}
(V_af)(x) = a(x)f(x)
\end{align}
is just the action of the Schwartz function $a$ on a spinor field (to be thought of 
as an element in $\mathcal{H}$) by pointwise multiplication.
On the other hand, if $a$ is interpreted as an element in $\mathscr{S}_\star(\mathbb{R}^4)$,
then the potential term $V_a$ takes e.g.\ the form
\begin{align} \label{eqn:nc_potential}
 (V_a f)(x) = a \star f (x) \,.
\end{align}
Now one can investigate the scattering problem of the Dirac equation with any of the potentials
$V_a$ of \eqref{eqn:comm_potential} or \eqref{eqn:nc_potential}. This, as we will see, renders
an action of $a$ on the field operators $\Psi(f)$. Let us recall how that proceeds for the potential
$V_a$ in the commutative case \eqref{eqn:comm_potential}: Here, one can put the field equation
\eqref{eqn:Dirac_lambda} in ``Hamiltonian form'' (or ``first order form''). That means, denoting
by $u_{\lambda,t}({\bf x}) = f_\lambda(t,{\bf x})$ the Cauchy-data at time $t$ of a solution $f_\lambda$ 
to \eqref{eqn:Dirac_lambda}, $f_\lambda$ is a solution iff the $u_{\lambda,t}$ satisfy a first-order
differential equation of the form
\begin{align} \label{eqn:first-order}
 \frac{d}{dt} u_{\lambda,t} + A_{\lambda,t}u_{\lambda,t} = 0 
\end{align}
where $A_{\lambda,t}$ is, at each $t$, a partial differential operator acting with respect to 
the spatial ${\bf x}$-coordinates (with $t$-dependent coefficients). In fact, a large class of
partial differential equations can be cast into this form, which often facilitates proving
existence and uniqueness of solutions to given Cauchy-data at some given value of time $t$.
Then one can define propagation operators $T_{\lambda,t} : u_{\lambda,0} \mapsto u_{\lambda,t}$ 
mapping data of a solution at time 0 to the data at time $t$.
Consequently, one can study the scattering problem in complete analogy to the scattering
problem in quantum mechanics, in first defining the M{\o}ller operators 
\begin{align}
 \Omega_{\lambda,\pm} = \lim_{t \to \pm\infty}\, T_{0,t}(T_{\lambda,t})^{-1}
\end{align}
and consequently, the scattering operator (here still at ``one-particle level''),
\begin{align}
 \boldsymbol{s}_{\lambda} = \Omega_{\lambda,+}(\Omega_{\lambda,-})^{-1} \,.
\end{align}
Under very general conditions, that scattering operator --- mapping a solution $f_0$ of the
Dirac equation $Df_0 = 0$ to another solution $\boldsymbol{s}_\lambda f_0$, i.e.\
$D\boldsymbol{s}_\lambda f_0 = 0$ --- induces a $C^*$-algebra morphism $\alpha_\lambda$ on the algebra
$\mathscr{F}(\boldsymbol{L}_0)$ of the quantized Dirac field on Minkowski spacetime by
\begin{align}
 \alpha_\lambda(\Phi(f_0)) = \Phi(\boldsymbol{s}_\lambda f_0)
\end{align}
with
\begin{align}
 \Phi(Rh) = \Psi(h) 
\end{align}
i.e.\ the $\Phi(f_0)$ are again algebraic generators of the CAR-algebra of the quantized Dirac field, but
labelled by solutions $f_0$ to the ``free'' Dirac equation $Df_0 = 0$, whereas the $\Psi(h)$ are labelled
by test-functions; the connection between these operators is a consequence of $\Psi(D h) = 0$.

Then one can differentiate $\alpha_\lambda(\Phi(f_0))$ with respect to the coupling strength
parameter $\lambda$, evaluated at $\lambda = 0$,
\begin{align} \label{eqn:deri-1}
\boldsymbol{\delta}_a(\Phi(f_0)) =  \left. \frac{d}{d\lambda} \right|_{\lambda = 0} \, \alpha_\lambda (\Phi(f_0)) = 
 \Phi(\boldsymbol{d}_a f_0)
\end{align}
where
\begin{align} \label{eqn:deri-2}
 \boldsymbol{d}_a f_0 = R V_af_0 \,,
\end{align}
i.e.\ pointwise multiplication of the solution $f_0$ to the free Dirac equation by $a$ followed by application of the
advanced-minus-retarded Green operator $R$ to produce again a solution to the free Dirac equation. One can check that 
\begin{align} \label{eqn:Bog-formula}
 \boldsymbol{\delta}_a(\Phi(f_0)) = i[:\Psi^+\Psi:\!{(a)},\Phi(f_0)]
\end{align}
where $:\Psi^+\Psi:$ is the normal-ordered squared Dirac field operator. (Here interpreted in the vacuum representation
of the ``free'', i.e.\ massless Dirac field; $\Psi^+$ is the Dirac-adjoint field to $\Psi$.) 
Thus, \eqref{eqn:Bog-formula} can be seen as an instance of ``Bogoliubov's formula'', deriving observable fields from differentiating
an $S$-matrix, or the corresponding scattering transformation (in our case, $\alpha_\lambda$) with respect to the
interaction strength. 

This provides already a hint on how one can expect the algebra $\mathcal{A}$ to make an appearance when setting up a quantum
field theory over a Lorentzian spectral triple: In the present case, that would occur via 
the operators ${:\Psi^+\Psi:}{(a)}$ which
are to be regarded as obervables of the quantized Dirac field (they correspond to the ``squared field strength'' weighted
with $a$ as smearing function).

Now we would like to implement a similar line of thought in the case of the quantized Dirac field on Moyal deformed
Minkowski spacetime and find counterparts to the ${:\Psi^+\Psi:}{(a)}$ in this case,
which then provides a handle on
how $\mathscr{S}_\star(\mathbb{R}^4)$ comes in play with the algebra $\mathscr{F}(\boldsymbol{L}_\star)$ --- and how this
interplay differes from the case of classical Minkowski spacetime. At that point, one encounters a difficulty: The potential
term $V_a$ of \eqref{eqn:nc_potential} is highly non-local, in particular, it is non-local in time.
This circumstance prevents turning the Dirac equation \eqref{eqn:Dirac_lambda} into a first-order system of the form \eqref{eqn:first-order}.
Nevertheless, as will be mentioned in Section 4, it is still possible to obtain (approximate) M{\o}ller operators and scattering operators for
solutions to \eqref{eqn:Dirac_lambda}, and thereby, implement ``Bogoliubov's formula''
also for the quantized Dirac field on Moyal deformed
Minkowski spacetime. We should like to point out that this approach of gaining observables of the quantized Dirac field
over some Lorentzian spectral triple is applicable generally  --- i.e.\ in principle beyond Moyal-like deformations
of Minkowski spacetime --- once one can define the scattering morphisms $\alpha_\lambda$ or their derivations 
$\boldsymbol{\delta}_a$ for the requisite $V_a$ corresponding to the Lorentzian spectral triple at hand. In fact, as we
will point out later, it works in similar fashion
for the localized Moyal-like deformations of Minkowski spacetime considered by Waldmann 
et al.\ \cite{BaWal,LechnerWaldmann:2011} 
although these non-commutative geometries have, at present,
not been described in terms of Lorentzian spectral triples.

\section{Wave equations with non-local $C_0^\infty$ kernel operators as potential terms}

In this section we will summarize the results of \cite{LechnerVerch} on the solution behaviour of
partial differential equations of the form
\begin{align} \label{eqn:Dlambda}
 D_\lambda f_\lambda = (D + \lambda W)f_\lambda = 0
 \end{align}
where $D$ is a normal hyperbolic operator, or pre-normal hyperbolic operator, and $f_\lambda$ is
in $C^\infty(\mathbb{R}^n,\mathbb{C}^N)$, $W$ is a $C_0^\infty$-kernel operator, and $\lambda$
is a complex parameter. The requiste definitions from \cite{LechnerVerch} are as follows:
\begin{defn}\label{definition:first}
	\begin{enumerate}
		\item A linear differential operator
		$D$ on $C^\infty(\mathbb{R}^n,\mathbb{C}^N)$ is called {\em normally hyperbolic} 
		if there exist smooth matrix-valued functions\\
		$U^0,...,U^s,V:\mathbb{R}^n\to\mathbb{C}^{N\times N}$ such that
		\begin{align}
			D
			=
			\frac{\partial^2}{\partial x_0^2}
			-
			\sum_{k=1}^s\frac{\partial^2}{\partial x_k^2}
			+
			\sum_{\mu=0}^s U^\mu(x)\frac{\partial}{\partial x_\mu}
			+V(x)
			\,.
		\end{align}
		\item A linear differential operator $D$ on $C^\infty(\mathbb{R}^n,\mathbb{C}^N)$ is 
		called {\em pre-normally hyperbolic} if $D$ is of first order, and there exists another first order
		differential operator $D'$ on $C^\infty(\mathbb{R}^n,\mathbb{C}^N)$ such that $D'D$ and
		 $DD'$ are normally hyperbolic.
		\item
		A $C_0^\infty$-kernel operator is a mapping $W:C^\infty(\mathbb{R}^n,\mathbb{C}^N)\to
		C^\infty(\mathbb{R}^n,\mathbb{C}^N)$ which can be represented as
	\begin{align}\label{eq:CinftyIntegralOperator}
		(Wf)(x)
		:=
		\int dy\,w(x,y)f(y)
		\,,\qquad f\in C^\infty(\mathbb{R}^n,\mathbb{C}^N),
	\end{align}
	where $w\in C_0^\infty(\mathbb{R}^n\times\mathbb{R}^n,\mathbb{C}^{N\times N})$. 
	The family of all $C_0^\infty$-kernel operators will be denoted by $\mathcal{W}$.
	\end{enumerate}
\end{defn}
As was already mentioned in the previous section, a normal hyperbolic or pre-normal hyperbolic operator
$D$ possesses unique advanced and retarded fundamental solutions (or Green operators) 
$R^\pm : C_0^\infty(\mathbb{R}^n,\mathbb{C}^N) \to C^\infty(\mathbb{R}^n,\mathbb{C}^N)$ which are 
characterized by the properties \eqref{eqn:Green-Op}. The main result of of \cite{LechnerVerch},
summarized in the subsequent Thm.\ \ref{Thm:2nd}, states
that this holds also for $D_\lambda = D + \lambda W$ where $D$ is (pre-) normal hyperbolic and 
$W \in \mathcal{W}$ provided that $|\lambda|$ is small enough, and even more generally. However, owing to the
non-local action of $W$ in general, the support properties of the Green-operators will not reflect the
causal propagation behaviour as in \eqref{eqn:Green-Op}. (For a graphical illustration, see \cite{LechnerVerch}.)
\begin{thm} \label{Thm:2nd}
Let $D_\lambda = D + \lambda W$ where $D$ is a normal hyperbolic or pre-normal hyperbolic operator on
$C^\infty(\mathbb{R}^n,\mathbb{C}^N)$, and $W \in \mathcal{W}$, $\lambda \in \mathbb{C}$. Suppose that
${\rm supp}(w) \subset K \times K$ for some compact subset $K$ of $\mathbb{R}^n$ where $w$ is the kernel
function of $W$. $R^\pm$ are the advanced/retarded Green operators of $D$.

	For sufficiently small $|\lambda|$, there are unique continuous linear 
	operators $R_\lambda^\pm:C_0^\infty(\mathbb{R}^n,\mathbb{C}^N)\to
	C^\infty(\mathbb{R}^n,\mathbb{C}^N)$ such that for any  $f,g\in C_0^\infty(\mathbb{R}^n,\mathbb{C}^N)$,
	the following relations hold:\footnote{In the present context, $\mathbb{R}^n$ is viewed as $n$-dimensional
	Minkowski spacetime with metric $(+,-,\ldots,-)$, and time-direction corresponding to increasing $x^0$-coordinate;
	$J^\pm(G)$ are the causal future/past sets of $G \subset \mathbb{R}^n$.}
	\begin{enumerate}
		\item $D_{\lambda}R^\pm_{\lambda}f=f=R^\pm_{\lambda}D_{\lambda}f$.
		\item ${\rm supp}(R^\pm_{\lambda} f) \subset 		J^\pm({\rm supp} f) \cup J^\pm (K)$.
		\item ${\rm supp}(R^\pm_\lambda f-R^\pm f)\subset J^\pm(K)$.
		\item If $J^\pm({\rm supp} f)\cap K=\emptyset$, then $R^\pm_\lambda f=R^\pm f$.
		\item If $D$ and $W$ are symmetric, i.e. $D=D^*$, $W=W^*$, and $\lambda\in\mathbb{R}$, then one has,
		for any $f,g \in C_0^\infty(\mathbb{R}^n,\mathbb{C}^N)$,
			\begin{align}
				\langle g, R_\lambda^\pm f\rangle
				=
				\langle R^{\mp}_\lambda g,f\rangle
				\,,
			\end{align}
			where the scalar product $\langle\,.\,,\,.\,\rangle$ is defined analogously as in
			\eqref{eqn:scalarproduct}.
Moreover, the dependence of $R^\pm_{\lambda}$ on $\lambda$ is meromorphic (in a suitable topology)\footnote{Thanks are 
due to Alexander Strohmaier for pointing this out} so that operators $R_\lambda^\pm$ with the stated
properties exist for all $\lambda$ values except
for a nonwhere dense set.
	\end{enumerate}
\end{thm}
For the proof, we refer to \cite{LechnerVerch}.

As mentioned already, since $W$ will in general act non-locally (in space and time), one cannot expect that the partial
differential equation \eqref{eqn:Dlambda} admits a well-posed Cauchy problem, even if $\lambda$ is chosen such that
the unique advanced and retarded Green operators $R^\pm_\lambda$ exist. We present two examples 
from \cite{LechnerVerch} illustrating such behaviour. The first example shows that one may construct a $W$ together
with $C_0^\infty$ initial data on a Cauchy surface such that there is no solution $f_\lambda$ to \eqref{eqn:Dlambda} with
the prescribed Cauchy data and $\lambda \neq 0$. 

\begin{prop}
	Let $D=\square$ be the d'Alembert operator, $\Sigma$ a Cauchy hyperplane, and $Wh:=\langle w_1,h\rangle w_2$ with $w_1,w_2\neq0$ 
	$C_0^\infty$ functions on $\mathbb{R}^n$
	such that ${\rm supp} w_1\subset\mathcal{O}_1$, ${\rm supp} w_2\subset \mathcal{O}_2$ with two spacelike separated double cones
	$\mathcal{O}_1$,
	$\mathcal{O}_2$ based on $\Sigma$.\footnote{That means that
	$\mathcal{O}_j = {\rm D}(\Sigma_j)$ where $\Sigma_j$ are
	open subsets of $\Sigma$ and ${\rm D}(\Sigma_j)$ denotes the open domain of dependence.}
	Pick Cauchy data $u$ on $\Sigma$ supported in $\mathcal{O}_1$ such that $f_0[u]$, 
	the unique solution to $Df_0[u] = 0$ with these Cauchy data,
	satisfies $\langle w_1,f_0[u]\rangle\neq0$, and also assume that $Rw_2\neq0$. 
	Then there exists no solution $f_\lambda$ to \eqref{eqn:Dlambda} with $\lambda \neq 0$ and Cauchy data~$u$.
\end{prop}

The next example from \cite{LechnerVerch} shows that solutions to \eqref{eqn:Dlambda} are in general not uniquely
determined by Cauchy-data.

\begin{prop}
	Let $D = \square$ be the d'Alembert operator, $W f=\langle w_1,f\rangle w_2$, with $w_1,w_2$ $C_0^\infty$ 
	functions on $\mathbb{R}^n$ having spacelike separated supports,
	and let $\Sigma$ be a Cauchy hyperplane such that ${\rm supp} w_2\subset \Sigma^-$, where
	$\Sigma^-$ denotes the open causal past of $\Sigma$ (excluding $\Sigma$).
	Denoting the Cauchy data of $Rw_2$ on $\Sigma$ by $u$, 
	let $f_\lambda:=f_0[u]-R^+w_2$. Then $w_1$ and $\lambda\neq0$ can be chosen in such a way that the $R_\lambda^\pm$ exist, 
	$f_\lambda$ is a non-zero solution to $D_\lambda f_\lambda = 0$, and $f_\lambda$ has zero Cauchy data on $\Sigma$.
\end{prop}
For proofs of these Propositions and graphical illustrations of the situations, we refer again to \cite{LechnerVerch}. 

Despite these difficulties, one can still show that there are uniquely determined M{\o}ller operators as well as scattering operators
as soon as the Green operators $R_\lambda^\pm$ exist. To describe the result obtained to this end in \cite{LechnerVerch}, we need to
collect some notation. Given $D$ and $W$ as before, we write $R_\lambda = R_\lambda^+ - R_\lambda^-$ and
${\rm Sol}_\lambda = R_\lambda C_0^\infty(\mathbb{R}^n,\mathbb{C}^N)$ for the space of solutions $f_\lambda$ to \eqref{eqn:Dlambda} obtained
from the Green operators (always assuming $|\lambda|$ sufficiently small so that $R_\lambda^\pm$ exist uniquely).  We usually write
$R = R_0$, in keeping with previously used notation. Again we assume that ${\rm supp}(w) \subset K \times K$ where $K$ is some
compact subset of $\mathbb{R}^n$, and we select two Cauchy-surfaces, $\Sigma_{\tau_\pm}$, such that $K$ is in the timelike past of
$\Sigma_{\tau_+}$ and in the timelike future of $\Sigma_{\tau_-}$. With these conventions, we can define the M{\o}ller operators
\begin{align}
     \Omega_{\lambda,\pm} : {\rm Sol}_\lambda \to {\rm Sol}_0 \, , \ \ \ \Omega_{\lambda,\pm}(R_\lambda h) = Rh \ \ \ ( \,h \in
     C_0^\infty(\Sigma^\pm_{\tau_{\pm}},\mathbb{C}^N) \,)
\end{align}
where $\Sigma^+_{\tau_+}$ is the open causal future of $\Sigma_{\tau_+}$ and $\Sigma^-_{\tau_-}$ is the open causal past
of $\Sigma_{\tau_-}$. Therefore, $\Omega_{\lambda,-}$ assigns to a solution $f_\lambda = R_\lambda h$ to  the ``interacting'' 
equation of motion $D_\lambda f_\lambda = 0$ the solution $f_0 = Rh$ to the ``free'' equation of motion $Df_0 = 0$ which
coincides with $f_\lambda$  everywhere sufficiently in the past of the interaction region $K$. This 
assignment is unambiguous
in view of Thm.\ \ref{Thm:2nd}, item 4. The action of $\Omega_{\lambda,+}$ is analogous, interchanging the past of the
interaction region $K$ with its future. Consequently, one can define the scattering operator
\begin{align}
 \boldsymbol{s}_\lambda : {\rm Sol}_0 \to {\rm Sol}_0\,,
 \ \ \ \boldsymbol{s}_\lambda = \Omega_{\lambda,+}(\Omega_{\lambda,-})^{-1}\,.
\end{align}

The following statement lists parts of the results in \cite{LechnerVerch}.

\begin{thm} ${}$ \\
\begin{itemize}
 \item[$(a)$] The M{\o}ller operators $\Omega_{\lambda,\pm}$ and the scattering operator 
 $\boldsymbol{s}_\lambda$ are linear bijections.
 \item[$(b)$] The scattering operator can be represented as
 $$ \boldsymbol{s}_\lambda =  1+RW\sum_{k=0}^\infty \lambda^{k+1}(- R^+W)^k $$
and the series converges in the operator norm on $L^2([-\tau,\tau] \times \mathbb{R}^{n-1},\mathbb{C}^N)$
where $[-\tau,\tau]$ is a finite but sufficiently large time interval. 
\item[$(c)$] For any $f_0\in {\rm Sol}_0$,
		\begin{align}
			\lambda \mapsto \boldsymbol{s}_\lambda f_0
		\end{align}
		is analytic in the nuclear topology of $C_0^\infty(\mathbb{R}^n,
		\mathbb{C}^N)$ on a finite disc around $\lambda=0$. In particular, given $f_0\in {\rm Sol}_0$, then
		\begin{align}
			\left.\frac{d (\boldsymbol{s}_\lambda f_0)}{d\lambda}\right|_{\lambda=0}
			=
			RWf_0
			\,.
		\end{align}
\end{itemize}

\end{thm}

\section{Discussion and Perspective}

One can show that the scattering operator $\boldsymbol{s}_\lambda$ induces, for suffiently small (real)
$\lambda$ a Bogoliubov transformation on the CAR algebra (if $D$ is Dirac operator) which provides
an abstract 2nd quantization of the Dirac field (i.e.\ of the solution space ${\rm Sol}_0$), as we
have sketched in Section 2. Similarly, $\boldsymbol{s}_\lambda$ induces a Bogoliubov transformation
on the CCR algebra which describes an abstract 2nd quantization of a Bosonic field in the case that
$D$ is a hyperbolic (wave-type) operator. Moreover, one can establish the relations
\eqref{eqn:deri-1} and \eqref{eqn:deri-2} for both the CAR and CCR quantized cases. For details,
see \cite{LechnerVerch}. 

There is also the following generalization: 
Suppose that there is an operator $W$ on $C^\infty(\mathbb{R}^n,\mathbb{C}^N)$ which
is no $C_0^\infty$ kernel operator, but such that
there is a sequence $W_\nu$, $\nu \in \mathbb{N}$, of $C_0^\infty$ kernel operators which 
approximate $W$ ``suitably''. 
That means, in particular, that (for the CAR/quantized
Dirac field case) the derivations
\begin{align}
 \boldsymbol{\delta}_{(\nu)}(\Phi(f_0)) = \left. \frac{d}{d\lambda}\right|_{\lambda = 0}
\alpha_{W_\nu,\lambda}(\Phi(f_0)) = \Phi(R W_\nu f_0)\,,
\end{align}
which are obtained from the scattering operators $\boldsymbol{s}_{W_\nu,\lambda}$ corresponding
to the potential terms $W_\nu$, should converge in the limit $\nu \to \infty$ to a derivation on
the algebra of field operators of the quantized Dirac field (in vacuum representation). 
That is in fact expected to hold when taking $W h = a \star h$ for a Schwartz function $a$ and
the Moyal product \eqref{MoyalProduct} with invertible $\theta$; it has been proved to hold
in the case that $\theta$ is degenerate and has no non-zero ``time-time'' components \cite{BorrisVerch}.
It is likewise expected to hold when taking $W h = a \circledast  h$ where 
$\circledast$ denotes a local non-commutative product introduced by Stephan Waldmann and co-authors
\cite{BaWal,LechnerWaldmann:2011}; see again \cite{LechnerVerch} for further discussion on this point.
In fact, this should be obtainable by standard arguments, and we hope to return to this issue
elsewhere. Assuming that convergence of $\boldsymbol{\delta}_{(\nu)}$ to a derivation 
$\boldsymbol{\delta}_W \equiv \boldsymbol{\delta}_a$ can be established in the mentioned cases,
one anticipates, as discussed in Section 2, that there is an assignment $a \mapsto X(a)$
of elements $a$ in the ``non-commutative spacetime algebra'' to quantum field operators
$X(a)$ that renders $\boldsymbol{\delta}_a(\Psi(h)) = 
i[X(a),\Psi(h)]$. The non-commutativity of the elements of the spacetime algebra is then reflected
in algebraic relations of the $X(a)$ for different $a$, e.g.\ in the behaviour of
the commutator $[X(a_1),X(a_2)]$. However, depending on the non-commutative
structure of the algebra $\mathcal{A}$ from which the $a_j$ are taken, it is likely that
useful properties of $[X(a_1),X(a_2)]$ could only be derived for special elements $a_1,a_2$
of $\mathcal{A}$. For instance, in case of $\mathcal{A} = \mathscr{S}_\star(\mathbb{R}^4)$,
the elements of the oscillator basis of the Moyal plane are promising candidates. This
should furthermore shed a light on potentially useful generalized locality concepts
for quantum field theories on non-commutative spacetimes, within the Lorentzian
spectral triple approach as well as beyond.

\subsection*{Acknowledgment}
Thanks are due to Alexander Strohmaier as indicated in a footnote in Section 2. Thanks are
also extended to the participants of the conference ``Quantum Mathematical Physics'', Regensburg, 2014, for 
remarks and comments. I would also like to thank the organizers and sponsoring institutions of that conference
for financial support.

\end{document}